\pdfoutput=1

\documentclass[%
 reprint, longbibliography,
superscriptaddress,
 amsmath,amssymb,
 aps,
 pra,
]{revtex4-2}

\usepackage[exponent-product = \cdot]{siunitx}
\DeclareSIUnit{\calorie}{cal}
\DeclareSIUnit{\Calorie}{\kilo\calorie}
\DeclareSIUnit{\atmass}{amu}
\usepackage{graphicx} %
\usepackage{bm}

\usepackage{hyperref}
\hypersetup{colorlinks=true,linkcolor=blue,citecolor=blue,urlcolor=blue}

\usepackage{color}
\newcommand{\red}[1]{{\color{black}{#1}}{}}

\usepackage{braket} %

\newcommand{\eu}{\mathrm{e}^}

\newcommand{\rmd}{\mathrm{d}}

\newcommand{\thalf}{{\ensuremath{\tfrac{1}{2}}}}

\providecommand{\mat}[1]{\mathsf{#1}}
\renewcommand{\mathbf}[1]{\bm{#1}}

\newcommand{\eqn}[1]{Eq.\,(\ref{#1})}

\newcommand{\Refx}[1]{Ref.~\citep{#1}}

\usepackage{dcolumn} %
\newcolumntype{d}[1]{D{.}{.}{#1}} %

\usepackage{multirow}
\usepackage{placeins}
\usepackage{booktabs}

\begin{document}

\title{Pattern stability in reaction--diffusion systems depends on path entropy}

\thanks{Supplemental material available}%
\author{Eric R. Heller}
\email{erheller@lbl.gov}
\affiliation{Department of Chemistry, University of California, Berkeley, 94720 Berkeley, USA}
\affiliation{Chemical Sciences and Materials Sciences Divisions, Lawrence Berkeley National Laboratory, Berkeley, CA, USA}
\author{David T. Limmer}
\email{dlimmer@berkeley.edu}
\affiliation{Department of Chemistry, University of California, Berkeley, 94720 Berkeley, USA}
\affiliation{Chemical Sciences and Materials Sciences Divisions, Lawrence Berkeley National Laboratory, Berkeley, CA, USA}
\affiliation{Kavli Energy NanoSciences Institute,
Berkeley, CA, USA}
\date{\today}

\begin{abstract}
\noindent
Reaction--diffusion systems driven far from thermodynamic equilibrium through the injection of energy can support multiple spatial patterns that persist as long-lived dynamical phases. The stability of these metastable phases is not determined by thermodynamics, but by the transition paths connecting them.
At finite particle numbers, intrinsic stochasticity induces rare transitions between competing patterns, rendering continuum mean-field descriptions insufficient, while exact stochastic simulations become computationally prohibitive in spatially extended systems. Here, we make absolute instanton rate calculations, including fluctuation and symmetry prefactors, tractable for transitions between metastable patterns in large nonequilibrium reaction--diffusion systems. Applied to a canonical Schlögl model and an experimentally motivated ATP-driven membrane kinase--phosphatase network, the resulting rates agree closely with reaction--diffusion master-equation benchmarks. We show that an effective path entropy can qualitatively alter stability at finite particle numbers by selectively enhancing one switching direction. These results establish path entropy as an organizing principle for finite-particle nonequilibrium pattern stability.

\end{abstract}

\maketitle

In equilibrium, the stability of a phase is determined by its free energy, which reflects a delicate balance between energetic and entropic contributions. Accordingly, the accurate prediction of equilibrium phase diagrams requires consideration of both quantities, as they jointly determine the complex structures that emerge \cite{ChandlerGreen,LubenskyBook}. Nonequilibrium steady states can exhibit similarly intricate behavior \cite{LimmerBook,Cates2015,Grafke2017}. Reaction--diffusion systems, for instance, sustain a wide variety of dynamic patterns whose organization can be tuned through external control parameters \cite{Turing1952,Cross1993}. Away from equilibrium, however, stability is no longer characterized by a static thermodynamic potential such as the free energy, but instead by dynamical properties like the lifetime of a metastable state \cite{Freidlin2012,Onsager1953,Graham1973}. 
The rates of transitions between metastable states depend sensitively on the ensemble of pathways connecting them, including both their likelihoods and their multiplicities. 
Using recently developed computational approaches based on nonequilibrium instanton theory \cite{NEQI,Weinan2004,Heymann2008,Grafke2015,Bouchet2016,Zakine2023,Lehmann2003}, we demonstrate how the entropy in path space, the diversity of transition pathways, can shape the phase behavior of reaction--diffusion systems.

Reaction--diffusion models are used across the physical and biological sciences because they capture a minimal but universal structure of local transformation and spatial transport. Most commonly, these models are formulated using deterministic partial differential equations \cite{Smoller1994}. 
\red{Within this framework, the stability of homogeneous steady states and the onset of Turing instabilities are identified through linear stability analysis, while propagating fronts, domain formation, and other nonlinear pattern dynamics are studied through the resulting reaction--diffusion equations, with broad applications in biological and nonequilibrium pattern formation \cite{Turing1952,Cross1993,Kondo2010,Halatek2018,Krause2021,karita2022scale}.}
While such descriptions capture typical spatiotemporal dynamics, they neglect the intrinsic stochasticity arising from the discrete and probabilistic nature of chemical reaction events, which can qualitatively alter phase stability, particularly through rare, noise-induced transitions between metastable states \cite{McQuarrie1967,LeeDeVille2006,Tian2006,Craciun2006,Smith2018,Lee2021}. 
Discrete stochastic descriptions based on the reaction--diffusion master equation account for these effects, but direct simulation of rare transitions becomes computationally prohibitive for spatially extended systems \cite{gillespie1976general,gillespie1977exact,gillespie2007stochastic,Nicholson2023}. 
\red{As a consequence, quantitatively resolving how intrinsic noise reshapes phase stability in spatially extended nonequilibrium reaction--diffusion systems remains difficult.}

\begin{figure}[t]
    \centering
    \includegraphics[width=1.0\linewidth]{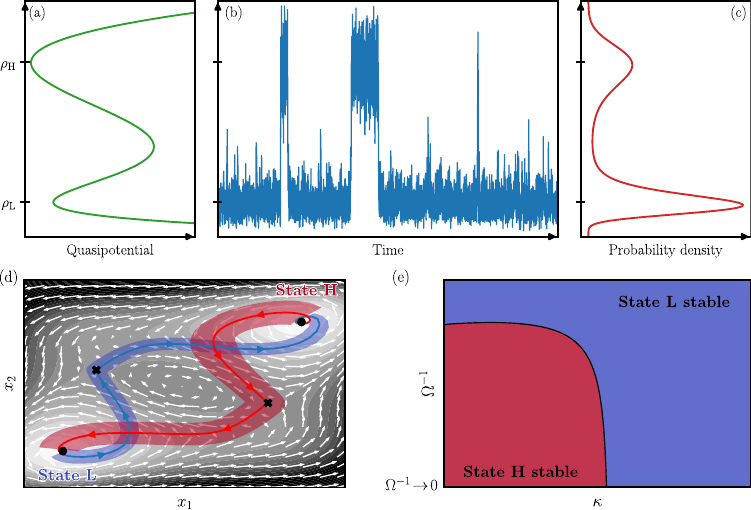}
    \caption{
    Overview of how transition path entropy can alter stability of metastable states in nonequilibrium systems.
    (a) Effective-action or quasipotential landscape along a one-dimensional order parameter, indicating that the high-density state ($\rho_\text{H}$) is favored in the absence of noise, $\Omega \rightarrow \infty$. 
    (b) A representative stochastic trajectory and (c) the corresponding steady-state probability density show that, at sufficiently strong noise, the low-concentration state ($\rho_\text{L}$) is more stable due to entropic enhancement of the $\rho_\text{H} \to \rho_\text{L}$ transition.
    (d) Two-dimensional representation of the system in concentration space, showing an underlying double-well potential (color map) together with a nonconservative force field (stream plot). 
    The forward (blue) and backward (red) optimal transition paths differ because detailed balance is broken. Broader fluctuations (transparent tubes) around the backward path indicate a larger statistical weight in path space, favoring it entropically at finite noise.
    (e) 
    Schematic nonequilibrium phase diagram as a function of a system parameter $\kappa$ and the noise strength $\Omega^{-1}$. In the zero-noise limit ($\Omega^{-1} \to 0$), stability follows from the action, whereas at finite noise, path entropy can qualitatively alter stability and eliminate the stability crossover altogether.
    }
    \label{fig:fig1}
\end{figure}

In order to bridge the gap between continuum mean-field descriptions and exact stochastic simulations, it is natural to focus on the rare transitions that control phase stability in noisy reaction--diffusion systems. 
In the weak-noise limit, these transitions are dominated by the optimal transition path or \textit{instanton}, allowing transition rates to be computed from a single trajectory. 
Instanton approaches therefore provide an efficient route to determining the stability of dynamical phases in nonequilibrium reaction--diffusion systems and can be used to predict phase stability by comparing instanton actions for forward and backward transitions between metastable states. 
Actions play a role analogous to energy barriers in equilibrium systems, but unlike equilibrium barriers, the actions for forward and backward transitions are not related by detailed balance \cite{Zakine2023,TanaseNicola2012}.
However, just as equilibrium phase stability depends not only on energy but also on entropy, the action alone does not fully determine nonequilibrium stability at finite noise \cite{Kikuchi2022}.
As illustrated schematically in Fig.~\ref{fig:fig1}, path entropy can be essential for predicting accurate nonequilibrium phase diagrams. 
\red{
By accounting for Gaussian fluctuations and translational symmetry modes around the optimal path within a numerically efficient framework \cite{NEQI,Bouchet2022,Grafke2023,Schorlepp2023a,Schorlepp2021}, 
we make full finite-noise rate calculations between competing patterns tractable for large nonequilibrium reaction--diffusion systems, allowing us to incorporate local path entropy into predictions of pattern stability. 
}

We first outline the instanton-based approach used to analyze rare transitions in nonequilibrium reaction--diffusion systems.
\red{In order to demonstrate how path entropy influences phase stability, we apply this framework to a canonical spatial Schlögl model \cite{kim2017stochastic} and to an ATP-driven membrane kinase–phosphatase network motivated by experiment \cite{Hansen2019}.}
Finally, we compare the path-based insights obtained for both systems to identify generic and system-specific features for system-size-dependent and particle-number-dependent stabilization of dynamic nonequilibrium phases.
\section*{Results and Discussion}
\subsection*{Instanton rates of reaction--diffusion systems}
We study reaction--diffusion systems in the limit where their dynamics are described by a chemical Langevin equation  \cite{Gillespie2000,Kampen2007}. Such a formulation assumes that the discrete Poissonian jump processes accompanying individual reactive events are well approximated by a Gaussian stochastic 
process in continuous state space. The chemical Langevin equation is valid when copy numbers $n^{(s)}$ of reactive molecular species $s$  within some correlation volume $\xi^d$ in $d$ dimensions are large but finite, and when many reactions occur over a characteristic correlation time $\tau$. The resultant equation of motion for a multi-component density field is
\begin{align}
    \label{equ:langevin}
    \dot{\rho}(\mat{x},t) = \mat{F}[\rho(\mat{x},t)] +D \nabla^2 \rho(\mat{x},t) +  \frac{\mathbf{\Lambda}[\rho(\mat{x},t)]}{\sqrt\Omega} \,\eta(\mat{x},t),
\end{align}
where $\rho(\mat{x},t)$ denotes the vector of continuous concentrations for each species $\rho^{(s)} = n^{(s)} / \Omega \xi^d$ in physical space $\mat{x}$, and $\eta$ is a vector of independent Gaussian white noises, one for each reaction or diffusion noise channel $r$, with zero mean and correlations $\braket{\eta_r(\mat{x},t) \eta_{r'}(\mat{x}',t')} = \delta_{r r'} \delta(\mat{x} - \mat{x}') \delta(t-t')$. 
The diffusion is conservative and ideal with diffusion constant $D$, the deterministic drift $\mat{F}$ describes the effects of reactions, and the noise matrix $\mathbf{\Lambda}$ encodes fluctuation measures for reaction and diffusion processes. %
The scale of the noise depends on the characteristic particle number $\Omega$ within a correlation volume $\xi^d$.
In the $\Omega \rightarrow \infty$ limit, the dynamics reduce to deterministic mass-action kinetics, whereas at finite noise, qualitatively different behavior can emerge, including rare transitions between metastable states.
\red{Because the Gaussianization intrinsic to the chemical Langevin approximation can fail to capture Poissonian rare-event tails even when typical fluctuations are accurately described \cite{Vellela2008,Schuettler2024}, we assess its validity separately through comparisons to direct simulations of the reaction--diffusion master equation and a channel-momentum diagnostic in the SI, finding excellent rate agreement for all systems considered.}
\begin{figure*}[t]
    \centering
    \includegraphics[width=0.8\linewidth]{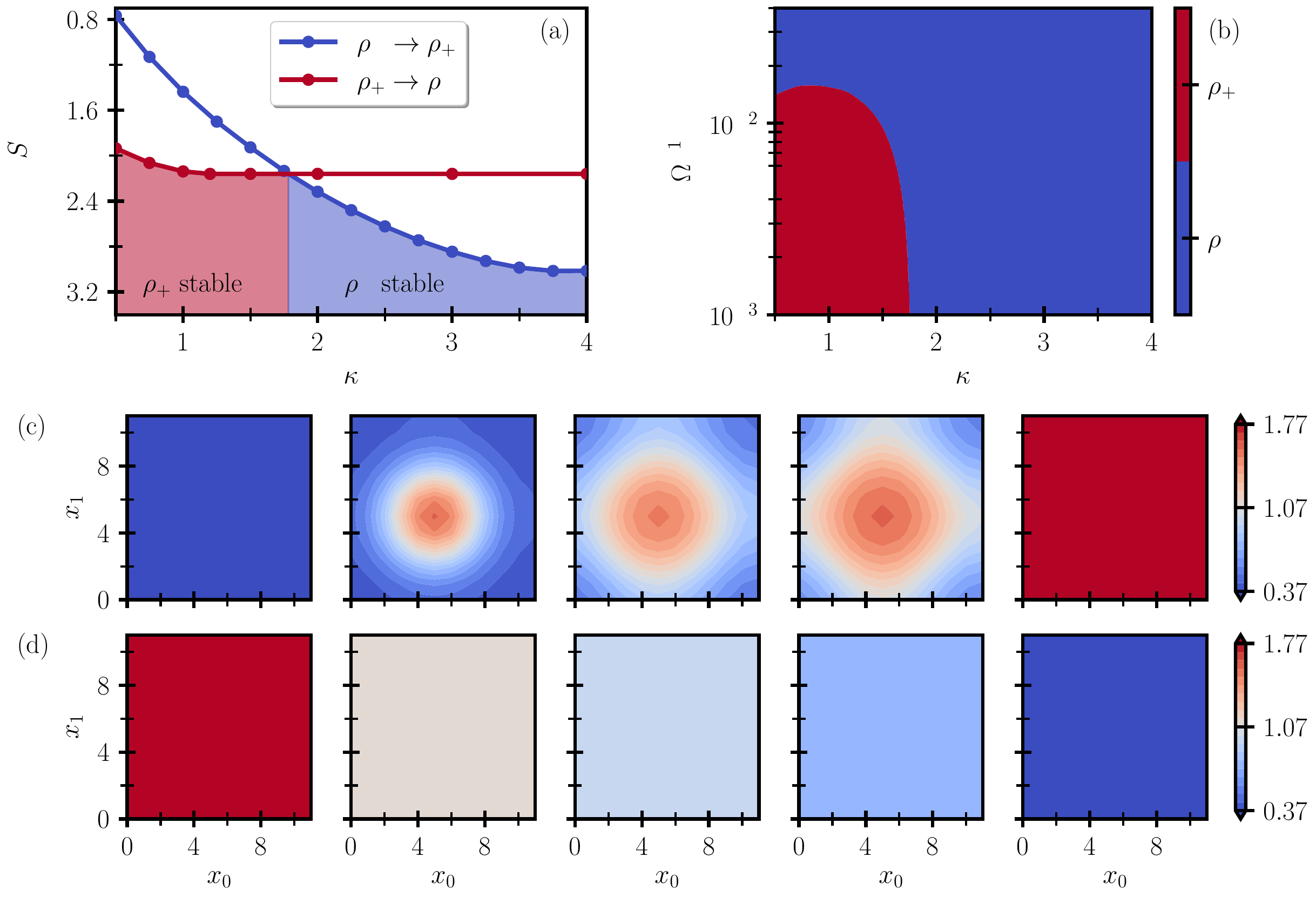}
    \caption{(a) Instanton actions for the forward (blue) and backward (red) transitions between the high- and low-concentration states of the 2$d$ Schl\"ogl model defined in the text.
    (b) Stability diagram, indicating the more stable phase as a function of $\kappa=D\tau/\xi^2$ and $\Omega$ based on the NEQI rate constants for the forward and backward transitions.
    Snapshots along the instantons of the forward (c) and backward (d) transitions at $\kappa=1.5$ with the color map indicating molecule concentration, where $\rho_{-}\xi^2 \approx 0.37$ and $\rho_{+}\xi^2 \approx 1.77$.
    The coordinates $x_0$ and $x_1$ span the two-dimensional domain in units of $\xi$. 
    \label{fig:2DSchloegl_stability}}
\end{figure*}

In order to characterize rare transitions between metastable states, we employ nonequilibrium instanton theory, which provides an asymptotic expression for transition rates in the weak-noise regime. Instanton theory identifies the optimal transition path between metastable states that minimizes the action
\begin{align}
    \label{equ:action}
   S &= \frac{1}{2} \int d\mat{x} \int \rmd t \,  [\dot{{\rho}} - \mat{v}({\rho})]^\text{T}  \boldsymbol{\chi}^{-1}({\rho}) \,[\dot{{\rho}} - \mat{v}({\rho})] ,
\end{align}
with $\mat{v}[{\rho}(\mat{x},t)] =\mat{F}[{\rho}(\mat{x},t)] + D \nabla^2 {\rho}(\mat{x},t)$ and $\boldsymbol{\chi}[\rho(\mat{x},t)] = \mathbf{\Lambda}[\rho(\mat{x},t)] \mathbf{\Lambda}^{\!\text{T}}[\rho(\mat{x},t)]$.
The resulting rate constant takes the asymptotic form
\begin{align}
    \label{equ:instrate}
    k &= A\, \ell^f \, \Omega^{f/2}  \, \eu{-\Omega S[\bar{\rho}]}, \qquad \left . \frac{\delta S}{\delta \rho} \right |_{\rho=\bar{\rho}} = 0 ,
\end{align}
where the action $S[\bar{\rho}]$ governs the exponential scaling of the rate and generalizes the notion of a barrier height in equilibrium systems. The prefactor $A$ captures fluctuations around the optimal path. Since the transition paths may involve the formation of interfaces with translational symmetry, we separately account for the effect of $f$ translational Goldstone modes with $\ell$ the linear system size in each spatial direction and $\ell^f$ as the measure of the associated symmetry manifold, area for $f$=2, length for $f$=1, or dimensionless for $f$=0. 
\red{The central technical advance of this work is the scalable evaluation of the full prefactor in \eqn{equ:instrate}, including spatial Goldstone modes, for high-dimensional nonequilibrium reaction--diffusion paths, enabling predictions of absolute transition rates and finite-noise stability diagrams beyond the action-only limit.}

Generically, the steady-state distribution of a nonequilibrium system is a complex function of its dynamical variables.  However, when the dynamics exhibit long-lived metastable states, the behavior can be described by coarse-grained transition rates between those states. The rates we quantify between metastable phases of reaction–diffusion systems in Eq.~\ref{equ:instrate} are precisely those that enter such a description. 
In this limit, the population of any metastable state is inversely proportional to the net rate at which it is exited.
By deducing a physically transparent form for these exit rates, we therefore obtain a means of understanding stability in nonequilibrium systems.
Most previous applications inferred stability from differences in the instanton action, corresponding to the exponential scaling limit \cite{Zakine2023}. For example, large instanton actions yield small exit rates and therefore more stable phases. Here, we extend this description by explicitly evaluating the fluctuation prefactor. 
This prefactor encodes an effective entropy in path space that can compete with the action at finite noise, as illustrated schematically in Fig.~\ref{fig:fig1}. 
Higher path entropy, generated either through large fluctuations around the optimal path or through a larger number of Goldstone modes, leads to larger exit rates and less stable phases.

In the following sections, we apply this framework to a spatial Schl\"ogl model and to a competing enzyme network, demonstrating how path entropy qualitatively modifies phase stability in driven reaction--diffusion systems. Both systems are considered in two dimensions in the main text, and additional simulations in one and zero dimensions are shown in the SI. To numerically evaluate the transition rates from the chemical Langevin description, we discretize space and time and employ our recent nonequilibrium instanton (NEQI) framework~\cite{NEQI}. 
Space is divided into $L$ voxels of length $\xi$, each containing a network of local chemical reactions, while adjacent compartments are coupled by diffusive particle hops with dimensionless microscopic rate $\kappa = D \tau/\xi^2$. 
Details of the implementation are presented in the SI, including a new computational scheme for evaluating the fluctuation prefactor that scales linearly with the number of time discretization points, enabling the treatment of complex spatially extended systems \red{requiring millions of space-time degrees of freedom.}
For the zero- and one-dimensional systems accessible to direct simulation, we validate the NEQI rates against the full stochastic dynamics generated by the reaction--diffusion master equation, as detailed in the SI. 
Unless stated otherwise, we work in unit systems where $\xi=\tau=1$.

\subsection*{Spatially extended Schl\"ogl model}
\label{sec:schloegl}

We first apply our approach to a spatially extended version of the classical Schl\"ogl model \cite{kim2017stochastic},
\begin{align}
    \notag
    \mathrm{A} \mathrel{\mathop{\rightleftarrows}^{\lambda_{+1}}_{\lambda_{-1}}} \mathrm{X} ,
    \qquad 
    2\mathrm{X} + \mathrm{B} \mathrel{\mathop{\rightleftarrows}^{\lambda_{+2}}_{\lambda_{-2}}} 3\mathrm{X} ,
\end{align}
where the concentrations of species A and B are held fixed by chemostats, and only a single density associated with species X evolves in time. For suitable reaction parameters, the deterministic drift of the corresponding chemical Langevin equation, $F(\rho) = \lambda_{+1} + \lambda_{+2} \rho^2 - \lambda_{-2} \rho^3 - \lambda_{-1} \rho$, exhibits bistability between a low-concentration state $\mat{\rho}_{-}$ and a high-concentration state $\mat{\rho}_{+}$.  Although the local drift can be written as the gradient of a double-well potential, the state-dependent noise violates the fluctuation--dissipation relation, and detailed balance is broken. As a consequence, the Schl\"ogl model is one of the simplest nonequilibrium chemical systems that exhibits bistability, making it a foundational model for understanding nonequilibrium chemical bistability.

The instanton actions for the forward and backward transitions between the two metastable states are shown in Fig.~\ref{fig:2DSchloegl_stability}(a) as a function of the microscopic diffusion constant.
For low $\kappa$, the high-concentration state is favored, whereas at large $\kappa$ the low-concentration state becomes more stable, with a crossover near $\kappa=D\tau/\xi^2 \approx 1.8$. 
This diffusion-constant-dependent crossover in stability is consistent with previous results for one-dimensional Schl\"ogl models \cite{TanaseNicola2012,Zakine2023}.
In the strong-diffusion limit, the actions plateau as the system behaves effectively as a single homogeneous compartment, whereas at $D=0$ each point in space evolves independently.
At intermediate values of $D$, however, the forward and backward transitions proceed via qualitatively different mechanisms. 
As shown in Fig.~\ref{fig:2DSchloegl_stability}(c--d), the forward transition from $\mat{\rho}_{-}$ to $\mat{\rho}_{+}$ nucleates a localized critical droplet, leading to a translational degeneracy of the instanton in two dimensions and two corresponding Goldstone modes, $f=2$. 
In contrast, the backward transition at the same $D$ follows a spatially homogeneous mechanism, $f=0$, and favors interface formation only at much slower diffusion.
These disparate mechanisms reflect the breaking of detailed balance and imply markedly different fluctuations around their optimal paths. 

Including the fluctuation prefactor allows us to compute absolute transition rates and construct the stability diagram shown in Fig.~\ref{fig:2DSchloegl_stability}(b) as a function of $D$ and the noise strength $\Omega^{-1}$. 
While the instanton action alone predicts a diffusion-dependent stability crossover, the inclusion of path entropy qualitatively modifies this picture. 
Below a characteristic particle number, $\Omega \lesssim 70$, the low-concentration state remains stable for all $D$, and the predicted crossover disappears. The disappearance of the action-based crossover at finite noise illustrates that stability in nonequilibrium reaction--diffusion systems cannot be inferred from the exponential action alone. 
Fluctuation effects in path space can offset substantial differences in action, leading to qualitatively different nonequilibrium phase diagrams. 
Note that the single-path approximation for the prefactor breaks down in a narrow regime in $D$ where competing mechanisms become comparable. 
In this crossover regime, we retain the optimized action and interpolate the logarithmic prefactor between neighboring converged calculations (see SI).
\subsection*{Competing enzyme network}
\label{sec:enzyme}

\begin{figure}[b]
    \centering
    \includegraphics[width=8.5cm]{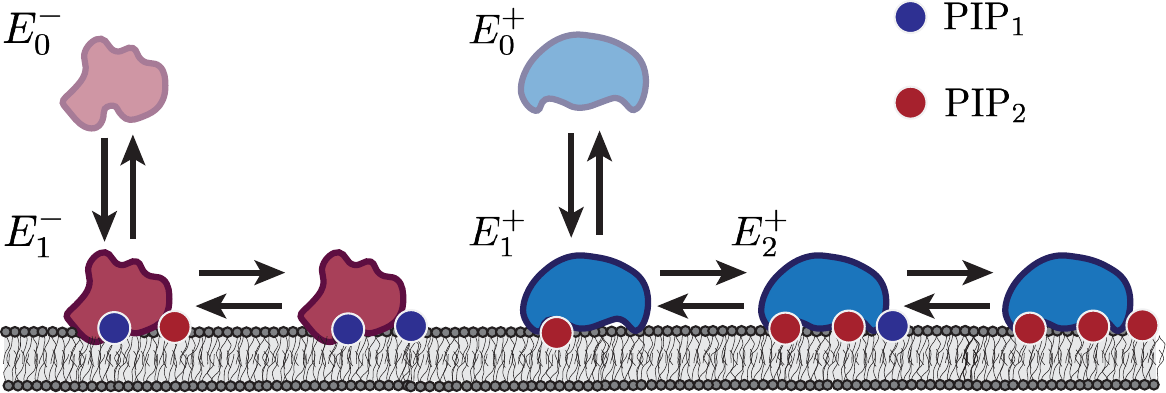}
    \caption{Schematic representation of the competing enzyme model studied in this work. The model consists of a lipid membrane containing PIP$_1$ and PIP$_2$ in contact with kinase and phosphatase enzymes in solution that can bind and catalyze lipid interconversion under ATP consumption. 
    }
    \label{fig:schematic}
\end{figure}
Building on the insights from the Schl\"ogl model, we next consider a more realistic nonequilibrium reaction--diffusion system inspired by the competitive kinase--phosphatase network studied experimentally in \Refx{Hansen2019}. 
In this system, the interconversion of the membrane lipids phosphatidylinositol-4-phosphate (PIP$_1$) and phosphatidylinositol-4,5-phosphate (PIP$_2$) is catalyzed by membrane-bound kinase and phosphatase enzymes.
Kinetic measurements identified the relevant species as the phosphatase in solution (E$_0^{-}$) and singly bound to PIP$_1$ (E$_1^{-}$), as well as the kinase in solution (E$_0^{+}$) and singly and doubly bound to PIP$_2$ (E$_1^{+}$, E$_2^{+}$) \cite{Hansen2019}. 
Under ATP consumption, the kinase (phosphatase) catalyzes phosphorylation (dephosphorylation) according to
\begin{align}
    \notag
    \mathrm{E}_1^{+} + \mathrm{PIP}_1 &\mathrel{\mathop{\rightarrow}^{\lambda_{6}\,\,\,}} \mathrm{E}_1^{+} + \mathrm{PIP}_2,
    \quad 
    \mathrm{E}_2^{+} + \mathrm{PIP}_1 \mathrel{\mathop{\rightarrow}^{\lambda_{7}\,\,\,}} \mathrm{E}_2^{+} + \mathrm{PIP}_2,\\
    \notag
    \mathrm{E}_1^{-} + \mathrm{PIP}_2 &\mathrel{\mathop{\rightarrow}^{\lambda_{8}\,\,\,}} \mathrm{E}_1^{-} + \mathrm{PIP}_1 ,
\end{align}
coupled to membrane binding and unbinding reactions, as illustrated in Fig.~\ref{fig:schematic}. 
The spatial model thus contains five relevant concentration fields subject to global conservation of the total lipid density.
\begin{figure*}[t]
    \centering
    \includegraphics[width=0.8\linewidth]{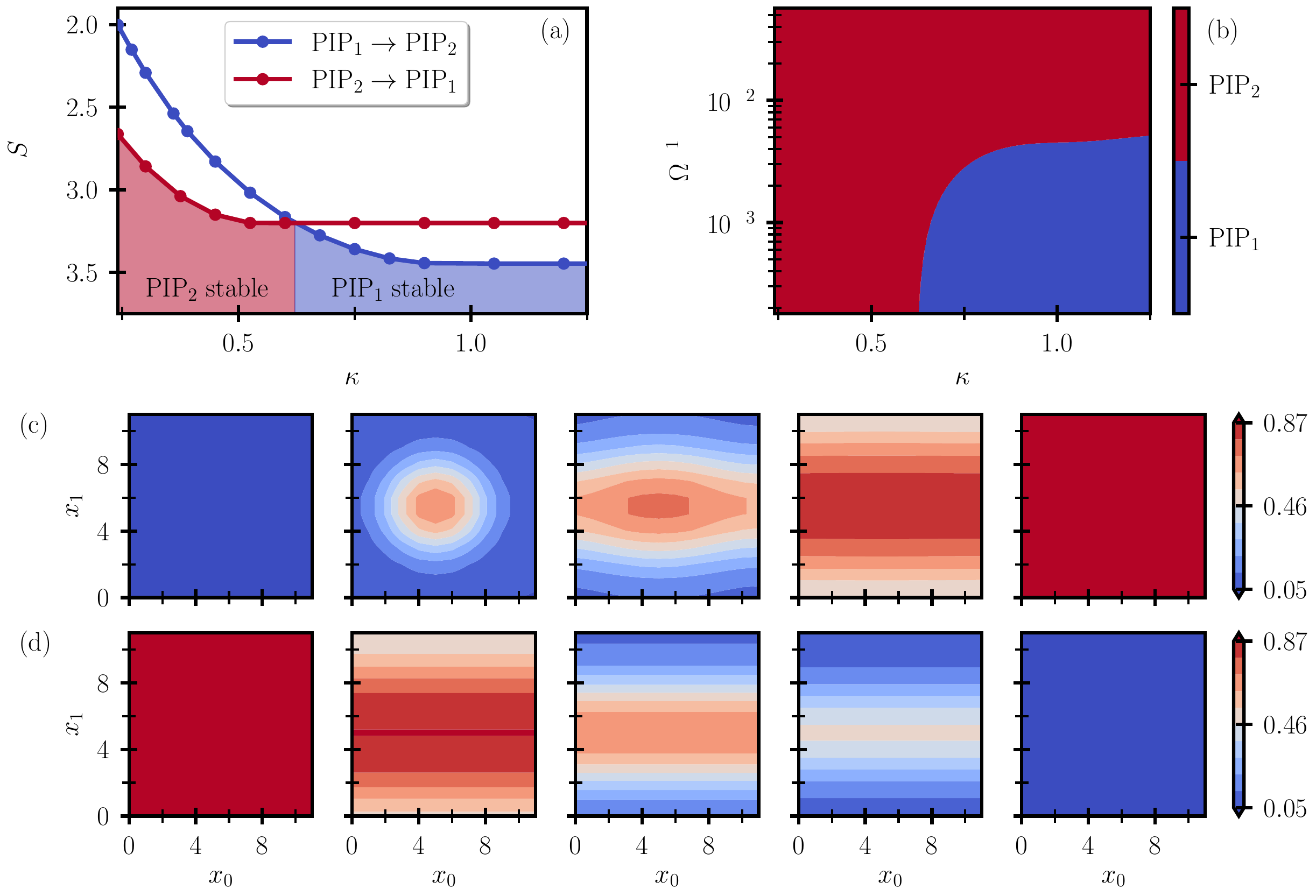}
    \caption{(a) Instanton actions for the forward (blue) and backward (red) transitions between the uniform PIP$_1$- and PIP$_2$-dominated states of the 2$d$ competitive enzyme model defined in the text.
    (b) Stability diagram, indicating the more stable phase as a function of $\kappa=D\tau/\xi^2$ and $\Omega$ based on the NEQI rate constants for the forward and backward transitions.
    Snapshots along the instantons of the forward (c) and backward (d) transitions at $\kappa=0.15$ with the color map indicating the PIP$_2$ concentration, where the PIP$_1$- and PIP$_2$-dominated states exhibit a PIP$_2$ concentration of $\rho^{\text{PIP}_2} \xi^2\approx 0.05$ and $\rho^{\text{PIP}_2} \xi^2\approx 0.87$.
    The coordinates $x_0$ and $x_1$ span the two-dimensional domain in units of $\xi$. 
    }
    \label{fig:2DEnzymeStability}
\end{figure*}

For a given set of rates and lipid density motivated by experimental observations, detailed in the Methods section,
this reaction network supports two metastable states dominated by either PIP$_1$ or PIP$_2$, consistent with observations showing macroscopic bistability in Ref.~\cite{Hansen2019}.
\red{The concrete objective is the prediction of which lipid state dominates at finite copy number.}
In contrast to the Schl\"ogl model, bistability here emerges from feedback between multiple interacting species.
High concentrations of PIP$_1$ or PIP$_2$ are directly coupled to increased populations of membrane-bound enzymes that catalyze the production of the respective lipid, reinforcing the dominant state. The enzyme system contains only first- and second-order reactions, leading to a drift in the reaction--diffusion equations that is at most quadratic in the species concentrations.
The bistability therefore does not arise from a double-well structure obtained by integrating the drift, as in the Schl\"ogl model,
but from genuinely nonequilibrium feedback between different chemical species encoded in a strongly nonconservative drift. 
The enzyme system is thus far from equilibrium both because it violates the fluctuation--dissipation relation and because its deterministic drift cannot be derived from a potential.

The phase stability in the spatially extended enzyme system in the limit of zero noise can be understood by optimizing instantons for the forward $\mathrm{PIP}_1 \rightarrow \mathrm{PIP}_2$ and backward $\mathrm{PIP}_2 \rightarrow \mathrm{PIP}_1$ transitions as a function of $D$. 
The resulting actions are shown in Fig.~\ref{fig:2DEnzymeStability}(a). 
As in the Schl\"ogl model, the exponential scaling predicts a diffusion-dependent crossover of the more stable state. 
In the strong-diffusion limit, both actions plateau and PIP$_1$ is favored, whereas at small $D$ the stronger diffusion dependence of the forward transition leads to an inversion of stability.

The corresponding instantons shown in Fig.~\ref{fig:2DEnzymeStability}(c--d) reveal that, in addition to uniform and droplet-forming paths, mechanisms with one-dimensional stripe-like interfaces appear due to finite system size, similar to mechanisms observed in driven nucleation problems \cite{Heymann2008a}. 
A species-resolved analysis (Fig.~SI.7) shows that the forward and backward transitions are initiated primarily by fluctuations in the membrane-bound enzyme species that dominates the respective product state, namely E$_2^{+}$ for $\text{PIP}_1 \to \text{PIP}_2$ and E$_1^{-}$ for $\text{PIP}_2 \to \text{PIP}_1$.
These distinct pathways are associated with markedly different fluctuation structures in path space, which are reflected in the stability diagram in Fig.~\ref{fig:2DEnzymeStability}(b).
As in the Schl\"ogl model, the crossover predicted from the action alone is recovered in the $\Omega \rightarrow \infty$ limit. 
At finite particle numbers, however, fluctuation effects qualitatively modify the phase diagram. For $\Omega \lesssim 100$, the PIP$_2$-dominated state becomes entropically favored over a broad range of diffusion constants.

As shown in Fig.~SI.10 of the SI, the NEQI+chemical Langevin approach provides rates in excellent agreement with master equation benchmarks.
The NEQI results thus provide a reliable estimate of the phase stability and the fluctuation-induced stabilization of the PIP$_2$ state in particular.
\subsection*{Role of path entropy in stability}
\label{sec:pathentropy}

The Schlögl model [Fig.~\ref{fig:2DSchloegl_stability}(b)] and the enzyme model [Fig.~\ref{fig:2DEnzymeStability}(b)] yield qualitatively similar stability diagrams, despite their very different reaction networks. 
In both cases, the exponential-scaling limit $\Omega \rightarrow \infty$ predicts a diffusion-dependent crossover of the most stable state. 
Identifying the PIP$_2$-dominated state as the high-concentration (and therefore high-noise) state for our parameters, we observe that in both systems the high-noise state is favored at slow diffusion, whereas the low-noise state dominates in the homogeneous limit. The origin of this diffusion-dependent stability crossover lies in the interplay between deterministic drift and state-dependent noise.
For the parameters considered, the drift alone favors the high-density state. 
At small $D$, the relevant saddle configuration is a small bubble of the high-density phase embedded in a low-density background.
Such a localized nucleus can form and grow more easily than the reverse process, which would require the creation of an extended low-density region within a high-density background. Consequently, the high-density state becomes more stable as $D$ decreases.
In the homogeneous $D \to \infty$ limit, however, spatial structure is suppressed.
As shown for the Schl\"ogl model in Fig.~SI.2 of the SI, the larger state-dependent noise amplitude modifies the effective quasipotential and overcompensates the drift advantage, stabilizing the low-density state overall. 
The diffusion-dependent crossover therefore reflects a shift between a bubble-nucleation regime dominated by the drift and a homogeneous limit dominated by the state-dependent noise.

Finite particle numbers modify this action-based picture of stability. In the limit of two metastable states, stability is controlled by rate ratios, not by action differences alone. 
It is therefore natural to define an effective nonequilibrium stability difference $\Delta \Phi(\Omega)=-\Omega^{-1} \ln{k_\text{fw}/k_\text{bw}}$ in terms of the forward and backward rate constants $k_\text{fw}$ and $k_\text{bw}$. Using our form of the rate [\eqn{equ:instrate}], this yields
\begin{multline}
    \Delta \Phi(\Omega)  = S_\text{fw} - S_\text{bw}\\
    -\Omega^{-1} \left[   \thalf (f_\text{fw} - f_\text{bw}) \ln{\Omega} + \ln\left(\frac{A_\text{fw}}{A_\text{bw}}\frac{\ell^{f_\mathrm{fw}}_\text{fw}}{\ell^{f_\mathrm{bw}}_\text{bw}}\right) \right],
\end{multline}
which makes explicit that the instanton action $S$ competes with entropic contributions from the prefactor. The prefactor is decomposed into contributions from translational symmetry modes and the fluctuation spectrum of all the remaining modes.
The diffusion-dependent crossover predicted from $S$ alone therefore becomes visible only once $\Omega$ is sufficiently large that action differences dominate over potentially large prefactor ratios.
At smaller $\Omega$, entropic advantages in path space can overwhelm the action difference and eliminate the crossover altogether.

The two models differ in which basin is entropically favored because the difference in path entropy between forward and backward transitions is dominated by different factors.
In the enzyme system, the $\text{PIP}_1 \rightarrow \text{PIP}_2$ transition strongly favors interface formation and exhibits soft interfacial modes that enhance the prefactor, entropically stabilizing the high-noise PIP$_2$ state. 
In the Schlögl model, by contrast, the low- and high-density basins differ more strongly in their intrinsic noise amplitudes. 
The larger effective noise in the high-density state softens fluctuations along the optimal paths in a way that entropically favors the high-to-low transition, thus entropically stabilizing the low-density state.

Overall, the comparison highlights a general principle for driven reaction--diffusion systems. Diffusion controls stability through the spatial geometry of the optimal bottleneck, bubble nucleation versus homogenized rearrangement, while finite particle numbers introduce path entropy as an additional stabilizing mechanism that depends on interfacial degeneracies, soft fluctuation modes, and state-dependent noise along the transition. Which effect dominates is system- and parameter-dependent and cannot be inferred reliably from the action or from geometry arguments alone. Quantitative predictions therefore require rate theories like NEQI to treat both the optimal paths and their fluctuations on equal footing.

\section*{Conclusions}

In summary, we have shown that phase stability in driven reaction--diffusion systems is governed not only by the optimal transition pathways between metastable states but also by the fluctuations around them.
Diffusivity and system size control whether rare transitions are dominated by bubble nucleation or homogeneous rearrangements.
In both the Schl\"ogl and enzyme models, the diffusion-dependent crossover in the most stable phase predicted from the instanton action alone emerges only in the weak-noise limit.
At finite particle numbers, entropic contributions associated with interfacial degeneracies and state-dependent noise can suppress or reverse this crossover. These results highlight path entropy as a central organizing principle for nonequilibrium phase stability. We suspect these effects are relevant for a broad range of systems in chemistry, molecular biology, and ecology where small spatial scales result in finite population sizes \cite{limmer2024molecular,birzu2018fluctuations,hagolani2019cell,hecht2010transient}.

\section*{Methods}
\label{sec:methods}
In order to study the chemical Langevin equation [\eqn{equ:langevin}] numerically, we explicitly partition space into $L$ correlation volumes of size $\xi^d$. Within voxel $l$, the concentration of species $s$ is $\rho_l^{(s)} = n^{(s)}_l / \Omega \xi^d$, where $n^{(s)}_l$ denotes the corresponding particle number. 
The chemical Langevin equation then reads
\begin{align}
    \label{equ:cle}
    \notag \dot{\rho}^{(s)}_l &= \sum_{r=1}^R 
    w_r^{(l)}(\rho)\, \nu_r^{(l,s)}  + \sum_{m\in nn(l)} \kappa^{(s)} (\rho_m^{(s)} - \rho_l^{(s)}) \\
    \notag
    &+ \Omega^{-1/2} \Bigg[
    \sum_{r=1}^R \sqrt{w_r^{(l)}(\rho)} \, \nu_r^{(l,s)} \eta_r^{(l)}(t)\\
    &+ \sum_{m\in nn(l)}  \left(
    \sqrt{ \kappa^{(s)} \rho_m^{(s)}} \eta^{(s)}_{m\to l}(t) 
    - \sqrt{ \kappa^{(s)} \rho_l^{(s)}} \eta^{(s)}_{l\to m}(t)
    \right)\Bigg] ,
\end{align}
where $\rho$ collects concentrations in all voxels and species, $w_r^{(l)}$
are reaction propensities, and $\nu_r^{(l,s)}$ are stoichiometric changes of species $s$ in voxel $l$ under reaction $r$. 
For elementary reactions, propensities follow mass-action kinetics and are given by products of reactant concentrations in voxel $l$ multiplied by the corresponding rate constants $\lambda_r$ \cite{Gillespie2000}.
Diffusion of species $s$ occurs between neighboring voxels $m\in nn(l)$ with microscopic diffusion constant $\kappa^{(s)}$.  The Gaussian noise variables have zero mean and correlations $\braket{\eta_r^{(l)}(t) \eta_{r'}^{(l')}(t')} = \xi^{-d} \, \delta_{r r'} \delta_{ll'} \delta(t-t')$, where $\xi^{-d}\, \delta_{ll'}$ represents the continuum noise correlation $\delta(\mat{x} - \mat{x}')$ in \eqn{equ:langevin} on a discrete voxel grid.
The first two terms in \eqn{equ:cle} describe deterministic reaction and diffusion, while the final term represents multiplicative reaction and diffusion noise obtained from approximating the Poissonian jump processes of the reaction--diffusion master equation in the large-$\Omega$ limit.
In the continuum notation of the main text, the deterministic drift and noise covariance in \eqn{equ:cle} recover $\mat{v}[{\rho}(\mat{x},t)] =\mat{F}[{\rho}(\mat{x},t)] + D \nabla^2 {\rho}(\mat{x},t)$ and $\boldsymbol{\chi}[\rho(\mat{x},t)] = \mathbf{\Lambda}[\rho(\mat{x},t)] \mathbf{\Lambda}^{\!\text{T}}[\rho(\mat{x},t)]$.
In the SI, we detail the conditions under which the chemical Langevin equation provides an accurate approximation of the master equation for rare reactive events.

For numerical evaluation, the path action [\eqn{equ:action}] is discretized into $N$ segments with time step $t_N = T/N$,
\begin{multline}
    \label{equ:SN}
    S_N(\mathbf{\rho}) = \frac{t_N}{2} \sum_{j=0}^{N-1} \xi^d \left[\frac{\mathbf{\rho}_{j+1} - \mathbf{\rho}_j}{t_N} - \mat{v}(\mathbf{\rho}_j)\right]^\text{T}\\  
    \times \mathbf{\chi}^{-1}(\mathbf{\rho}_j) \left[\frac{\mathbf{\rho}_{j+1} - \mathbf{\rho}_j}{t_N} - \mat{v}(\mathbf{\rho}_j)\right] \, ,
\end{multline}
where, $\mathbf{\rho}$ contains all system copies or \textit{beads}, $\mathbf{\rho}_j$, along the path with endpoints fixed at the reactant fixed point and the unstable saddle separating the metastable states. 
Each bead configuration is a vector containing the density in every correlation volume. The quadratic form in \eqn{equ:SN} therefore corresponds to evaluating the spatial integral in \eqn{equ:action} as a discrete sum.
The instanton is obtained by minimizing this action with respect to the intermediate bead configurations $\mathbf{\rho}_1,\ldots,\mathbf{\rho}_{N-1}$ using Newton--Raphson-type optimizers \cite{NEQI,InstReview,PhilTransA}.  
The results must be converged with respect to $T$ and $N$, which can be sped up by allowing variable time steps \cite{NEQI} or transforming to a path-length coordinate \cite{Heymann2008}. 

From the instanton, the rate constant is evaluated as \cite{NEQI}
\begin{align}
    \label{equ:kneqi}
    k &\sim \frac{\zeta}{2\pi t_N}
    \sqrt{\frac{B_t \xi^{2d}}{t_N p_\text{f}^2 \Sigma}} \, 
    \eu{-\Omega S_N},
\end{align}
where $\Sigma = \det'[t_N \nabla^2_N S_N / \xi^d] \prod_{j=1}^{N}\text{det}[\mathbf{\chi}(\mathbf{\rho}_j)]$
encodes fluctuations around the optimal path, $\nabla_N = (\partial/\partial \mathbf{\rho}_1,\ldots, \partial/\partial \mathbf{\rho}_N)$ represents the gradient with respect to bead coordinates, 
$\nabla^2_N S_N$ is the Hessian of the discretized action,
and the prime indicates omission of zero eigenvalues from temporal and spatial symmetries as well as the projected reaction coordinate. 
The magnitude of the endpoint momentum is denoted by $p_\text{f}$, and the Jacobians  $B_t$ and $\zeta$ arise from integration over zero modes and are detailed in the SI. 
The instanton results were converged by systematically increasing the number of beads $N$ until doubling $N$ changed the action by less than 1\% and the prefactor by less than 20\%, requiring between $2000$ and $12000$ beads depending on system parameters. 
Convergence with respect to the total propagation time $T$ was verified by repeating the calculations at increasing $T$ and by confirming a numerically vanishing eigenvalue associated with time-translation symmetry.  

The two-dimensional Schl\"ogl and enzyme models are defined on a $12\times 12$ grid with periodic boundary conditions. The reaction rates of the Schl\"ogl model are chosen as $\lambda_{+1} \tau=0.73$, $\lambda_{+2}\tau=3.7$, $\lambda_{-1}\tau=3.2$,  $\lambda_{-2}\tau=1.2$.
In addition to catalytic lipid interconversion reactions
\begin{align}
    \notag
    \mathrm{E}_1^{+} + \mathrm{PIP}_1 &\mathrel{\mathop{\rightarrow}^{\lambda_{6}\,\,\,}} \mathrm{E}_1^{+} + \mathrm{PIP}_2,
    \quad 
    \mathrm{E}_2^{+} + \mathrm{PIP}_1 \mathrel{\mathop{\rightarrow}^{\lambda_{7}\,\,\,}} \mathrm{E}_2^{+} + \mathrm{PIP}_2,\\
    \notag
    \mathrm{E}_1^{-} + \mathrm{PIP}_2 &\mathrel{\mathop{\rightarrow}^{\lambda_{8}\,\,\,}} \mathrm{E}_1^{-} + \mathrm{PIP}_1 ,
\end{align}
the enzyme model includes membrane binding and unbinding reactions
\begin{align}
    \notag
    \mathrm{E}_0^{+} + \mathrm{PIP}_2 &\mathrel{\mathop{\rightarrow}^{\lambda_{0}\,\,\,}} \mathrm{E}_1^{+} + \mathrm{PIP}_2,
    \quad 
    \mathrm{E}_1^{+} + \mathrm{PIP}_2 \mathrel{\mathop{\rightarrow}^{\lambda_{1}\,\,\,}} \mathrm{E}_2^{+} + \mathrm{PIP}_2,\\
    \notag
    \mathrm{E}_0^{-} + \mathrm{PIP}_1 &\mathrel{\mathop{\rightarrow}^{\lambda_{4}\,\,\,}} \mathrm{E}_1^{-} + \mathrm{PIP}_1,  \\
    \notag
    \mathrm{E}_1^{+} &\mathrel{\mathop{\rightarrow}^{\lambda_{2}\,\,\,}} \mathrm{E}_0^{+}, 
    \quad
    \mathrm{E}_2^{+} \mathrel{\mathop{\rightarrow}^{\lambda_{3}\,\,\,}} \mathrm{E}_1^{+}, 
    \quad
    \mathrm{E}_1^{-} \mathrel{\mathop{\rightarrow}^{\lambda_{5}\,\,\,}} \mathrm{E}_0^{-} ,
\end{align}
as well as weak non-specific binding reactions introduced to ensure ergodicity 
\begin{align}
    \notag
    \mathrm{E}_0^{+} + \mathrm{PIP}_1 \mathrel{\mathop{\rightarrow}^{\lambda_{9}\,\,\,}} \mathrm{E}_1^{+} + \mathrm{PIP}_1,
    \quad
    \mathrm{E}_0^{-} + \mathrm{PIP}_2 \mathrel{\mathop{\rightarrow}^{\lambda_{10}\,\,\,}} \mathrm{E}_1^{-} + \mathrm{PIP}_2 .
\end{align}
Throughout, we choose the following rates for the first-order reactions
$
\lambda_0 \tau= 1, 
\lambda_2 \tau= 4,
\lambda_3 \tau= 0.1,
\lambda_4 \tau= 0.05,
\lambda_5 \tau= 0.2,
\lambda_9 = 0.1\lambda_0,
\lambda_{10}= 0.1\lambda_4
$,
and the second-order reactions
$
\lambda_1 \tau= 0.1635, 
\lambda_6 \tau= 0.556, 
\lambda_7 = \lambda_8 = 4\lambda_6, 
$
where the enzyme concentrations in solution (E$_0^{+}$, E$_0^{-}$) are absorbed into pseudo-first-order rate constants.
The total lipid number is conserved, and the total lipid density is fixed at $\rho_\text{tot} \xi^2= 1$.
\\

\noindent
\textbf{Data availability:}
The source code for an instanton calculation, the data that support the findings of this study, and several videos of time evolutions of the different systems in this work along the instantons are openly available in the SI and online \cite{SourceData}.
\\

\noindent
\textbf{Acknowledgments:}
ERH is grateful for financial support from the Swiss National Science Foundation through Grant 214242. DTL was supported by the Condensed Phase
and Interfacial Molecular Science Program (CPIMS)
of the U.S. Department of Energy under contract no.
DEAC02-05CH11231. This research used the Lawrencium computational cluster resource provided by the IT Division at Lawrence Berkeley National Laboratory under the same contract.

\bibliography{references,other}
\end{document}